# Evaluation of cross section of elastic scattering for non-relativistic and relativistic particles by means of fundamental scattering formulas

Huai-Yu Wang

Department of Physics, Tsinghua University, Beijing, 100084 China

wanghuaiyu@mail.tsinghua.edu.cn

**Abstract**

In evaluating differential cross section of elastic scattering, different theories were applied to low-momentum and relativistic particles. For low-momentum motion, Lippmann-Schwinger scattering equation was applied, called fundamental formula; while for relativistic particles, a general scattering theory was used which calculates $S$ matrix. In this paper, Lippmann-Schwinger equation is applied uniformly to both low-momentum and relativistic particles. The cross sections are valuated to the first order of Born approximation. One-body time-independent Green's functions for relativistic free particles are given. Compared to the general scattering theory, the fundamental theory has a clearer physical picture and the approximations made are more explicit.

## 1 Introduction

Relativistic quantum mechanics equations (RQMEs) have both positive kinetic energy (PKE) and negative kinetic energy (NKE) solutions. The latter is usually explained as antiparticles. The author thinks that the NKE solutions do not means antiparticles, but simply are particles with NKE [1]. In the author's opinion, the NKE and PKE solutions ought to be treated on an equal footing. Hence, all the field concerning the PKE ought to be studied in the aspect of the NKE. Based on this premise, a series work has been done [1-7]. Up to now, the research is within the scope of relativistic quantum mechanics (RQM). The author believes that a RQME is a self-consistent theory. Therefore, any problem should be figured out within the RQM in which it is yielded. One remarkable example was the famous Klein's paradox [2]. After all the problems in the RQM are clear, the author will enter quantum electrodynamics (QED).

Historically, quantum mechanics (QM), including the RQMEs, emerged first. Then, second quantization of Schrödinger equation and RQMEs was implemented, and QED was developed. Later quantum field theory (QFT) was established. Theories develop from the lower levels to higher ones gradually. The RQM is the first field the author planned to inspect, and next will be quantum electrodynamics.

In the QED, one mainly deals with scattering problems, and calculates the cross sections. The author notices that the scattering theories in QED and in QM have different formulas. Therefore, before using the scattering theory in QED, the scattering formulas in QM should be made clear.

This paper discusses the evaluation of differential cross section of elastic scattering of three kinds of particles. The three kinds of particles refer to low-momentum particles observing Schrödinger equation and relativistic particles with spins zero and 1/2. They are respectively called Schrödinger particles, spin-0 particles and Dirac particles.

Although the methods evaluating the cross section have been sophisticated, it seems that different methods, as well as different concepts, were applied to low-momentum and relativistic particles.

For low-momentum motion, if an incident particle meets a potential centered at the origin, then after it is scattered, the outgoing wave, the final state $\psi$, is composed of two terms [8,9,10]. One is along the incident direction, and the other is scattered one divergent to all directions in space. The rigorous formula of this theory is well-known Lippmann-Schwinger equation [11]. This equation was derived from time-independent Schrödinger equation. From an observer away from the scattering center, the scattered wave behaves as a spherical one, although not necessarily isotropic. So, this wave can be written as a spherical wave multiplied by a factor called scattering amplitude. From the scattering amplitude, the differential cross section can be evaluated. This theory is called fundamental scattering theory.

However, for relativistic particles, a routine, called general scattering theory [12] was employed. This is to utilize $S$ matrix. The scattering matrix element $S_{fi}$ is defined as the probability amplitude for the transition $\Psi_i \to \Psi_f$, where $\Psi_i$ denotes the free initial state long before and $\Psi_f$ the free final state long after the scattering. In other terms, $S_{fi}$ is the projection of $\psi_i$ onto $\Psi_f$, where $\psi_i$ is the evolutionary state during the scattering process originating from $\Psi_i$ [13]. This general theory, in the author's opinion, makes the evaluation of the cross section complicated. Here the so-called evolution state $\psi_i$ is actually the final state $\psi$ in the fundamental scattering theory. That is to say, the concept of the final state is different from that in the fundamental scattering theory. In this scattering theory, Hamiltonian is time-dependent.

Personally, RQMEs are applicable to all momenta, including very low ones. Indeed, the low-momentum approximations of the RQMEs lead to Schrödinger equation. Therefore, when the general theory is applied to low-momentum motion, it is just the fundamental scattering theory. On the contrary, the fundamental scattering theory should be, at least in some extent, applicable to RQMEs.

In fact, Mott [14] and others calculated the cross section of Dirac particles long before the emergence of the general scattering theory. At that time, there were no concepts involved in QFT such as interaction picture, time evolution operator and $S$ matrix. This fact demonstrates that at least to some approximation the cross section of relativistic particles can be evaluated without resorting to the general scattering theory.

The fundamental scattering theory could also be recast into the form the same as the general theory [13], but that was not necessary.

The present work shows that to the first order of Born approximation, the fundamental theory can be used to calculate the cross sections of all the three kinds of particles.

Section 2 will briefly review the fundamental theory including Lippmann-Schwinger equation and the formula of cross section. This theory makes use of time-independent Green's functions of free particles. We will also mention the concept of $S$ matrix in the general theory, illustrating the discrepancy between the two theories. Section 3 evaluates the cross sections of the first order of Born approximation for all the three kinds of particles in three-dimensional space. Section 4 is our conclusions. Appendix A presents the Green's functions of all the three kinds of free particles in one-, two- and three-dimensional spaces. Appendix B calculates the cross sections of the three kinds of particles in one- and two dimensional spaces, demonstrating the universal applicability of the fundamental theory.

**2 Fundamental Formulas of Elastic Scattering**

In this work, we merely consider time-independent scattering.

The Hamiltonian of a free particle is denoted by $H_0$. Its energy spectrum is denoted by $E_k$ where $k$ means wave vectors. The spectrum is real and continuous, and the corresponding eigen wave functions are denoted by $\varphi_k$.

The free incident particle is of the form of plane wave

$$\varphi_0(\mathbf{r}) \propto e^{i\mathbf{p}\cdot\mathbf{r}}. \tag{2.1}$$

Its eigen equation is

$$H_0 \varphi_0(\mathbf{r}) = E^+ \varphi_0(\mathbf{r}). \tag{2.2}$$

Here $\varphi_0(\mathbf{r})$ is one of $\varphi_k$ and $E$ is one of $E_k$. The $E$ is added an infinitely small imaginary part, $E^+ = E + i0^+$. The corresponding Green's function $G_0(\mathbf{r},\mathbf{r}',z)$ satisfies the following equation

$$(z - H_0) G_0(\mathbf{r},\mathbf{r}',z) = \delta(\mathbf{r} - \mathbf{r}'), \tag{2.3}$$

where $z$ is a complex parameter. The Green's functions are presented in appendix A. They can be computed by eigen wave functions [15,16]. Now that the spectrum $E$ is real and continuous on the real axis, the Green's function is not defined on the real axis. Nevertheless, on can define a side limit of the Green's function as follows.

$$G_0^+(\mathbf{r},\mathbf{r}',E) = G_0(\mathbf{r},\mathbf{r}',E+\mathrm{i}0^+).\tag{2.4}$$

Scattering center fixed at the origin generates a potential presented by Hamiltonian $H_1$. The total Hamiltonian is

$$H = H_0 + H_1.\tag{2.5}$$

After scattering, the wave is denoted by $\psi$. It is the eigen function of $H$. We only consider elastic scattering so that the particle's energy $E$ remains unchanged. After the scattering, the eigen equation is

$$H\psi(\mathbf{r}) = E^+\psi(\mathbf{r}).\tag{2.6}$$

It results from the above equations that

$$\psi(\mathbf{r}) = \varphi_0(\mathbf{r}) + \int \mathrm{d}\mathbf{r}' G_0^+(\mathbf{r},\mathbf{r}';E_i)H_1(\mathbf{r}')\psi(\mathbf{r}'),\tag{2.7}$$

where Eq. (2.4) is employed. This is Lippmann-Schwinger equation [11]. They started from time-dependent formulas to obtain this equation through a variational procedure. In the present case, the Hamiltonian (2.5) is independent of time, so that we acquire Eq. (2.7) from the above equations directly. This is a concise way [17,18]. One can also start from the time-dependent formulas and makes use of the feature that Hamiltonian does not rely on time, so as to gain Eq. (2.7) [8].

Here we stress that although in textbooks Eq. (2.7) is derived for low-momentum particles, from the above procedure, the form of Hamiltonian (2.5) is not constrained. Therefore, Eq. (2.7) should be applicable to relativistic particles either.

Up to this step, no approximation is made in obtaining Eq. (2.7). Nevertheless, it is a self-consistent equation. The right hand side contains the wave function after scattering. Therefore, to carry out calculation, approximations have to be made.

Now let us make Born approximation. Iterating the equation once leads to

$$\psi(\mathbf{r}) = \varphi_0(\mathbf{r}) + \int \mathrm{d}\mathbf{r}' G_0^+(\mathbf{r},\mathbf{r}';E_i)H_1(\mathbf{r}')\varphi_0(\mathbf{r}').\tag{2.8}$$

The concrete form of Green's function $G_0^+$ depends on Hamiltonian.

Equation (2.8) is rewritten as

$$\psi(\mathbf{r}) = \varphi_0(\mathbf{r}) + \varphi_s(\mathbf{r}),\tag{2.9}$$

where $\varphi_s$ is the second term of (2.8), called scattering wave.

The potential $H_1$ generated by the scattering center plays a role within a finite region near the origin and decays from the center rapidly enough. In considering scattering problem, the observer is always distant from the origin, where the potential is so weak that it can be neglected. At this distance, the scattering term can be rewritten

in the form of a spherical wave

$$\varphi_s(\boldsymbol{r}) = f(\theta,\varphi)\frac{e^{ikr}}{r}. \tag{2.10}$$

The factor $f(\theta,\varphi)$ is called scattering amplitude. The key to obtain the form of (2.10) from (2.8) through appropriate approximations depends on the form of Green's function.

In Fig. 1 sketched is the physical picture of the scattering based on Eqs. (2.8)-(2.10). There was a more detailed picture in [10]. In the following we always assume the incident particle moves along the z direction.

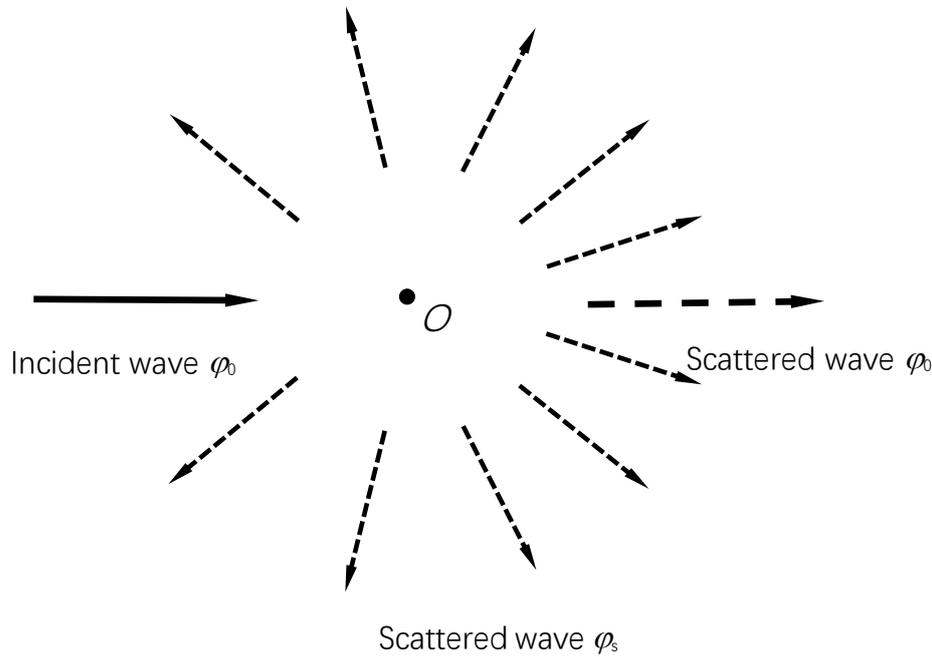

Incident wave $\varphi_0$

Scattered wave $\varphi_0$

Scattered wave $\varphi_s$

**Figure 1**. The solid line means the incident wave $\varphi_0$. The dashed lines mean the waves after scattering, which are composed of two parts: that exactly the same as the incident wave $\varphi_0$ and scattered wave $\varphi_s$.

As the wave function $\psi$, Eq. (2.9), is substituted into the expression of probability current density, the obtained $\boldsymbol{j}(\psi)$ usually contains three terms:

$$\boldsymbol{j}(\psi) = \boldsymbol{j}_0(\varphi_0) + \boldsymbol{j}_s(\varphi_s) + \boldsymbol{j}_c(\varphi_0,\varphi_s). \tag{2.11}$$

The first term $\boldsymbol{j}_0$ is the incident current density which is along the z direction; the second term $\boldsymbol{j}_s$ is the scattering current density, which depends on azimuth angle $(\theta,\varphi)$; the last term $\boldsymbol{j}_c$ comes from the interference between $\varphi_0$ and $\varphi_s$. The forms of $\varphi_0$ and $\varphi_s$ are respectively (2.1) and (2.10). Here we are discussing elastic scattering, so that the values of the wave vectors in (2.1) and (2.10) are the same, $p = k$. Thus, in $j_c$ there is necessarily a factor [12,19]

$$e^{-i\mathbf{p}\cdot\mathbf{r}}e^{ikr} = e^{-ipr\cos\theta}e^{ikr} = e^{ikr(1-\cos\theta)} \qquad (2.12)$$

or its complex conjugate. This factor oscillates rapidly as the distance $r$ increases, so that it has no observable effect to a distant instrument, and can be neglected [12,19].

We recall the formula evaluating the differential cross section. The cross section had an explicit definition [8,20,21]. A concise form of the formula is that [21]

$$\sigma(\theta,\varphi) = \frac{r^2 j_s}{j_0}. \qquad (2.13)$$

Although most textbooks do not adopt this expression, the author thinks that it is most convenient, as will be shown below.

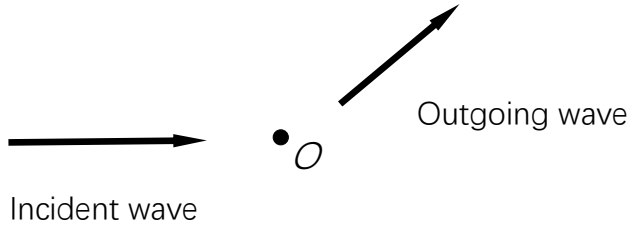

**Figure 2**. The physical picture of elastic scattering usually seen in literature [21].

Here we briefly mention the general scattering theory used for relativistic particles. The key is to calculate $S$ matrix element between initial and final states. The states are respectively plane waves long before and after the scattering. The corresponding picture is illustrated in Fig. 2, which is somehow similar to classical scattering.

The explicit definition of the mentioned $S$ matrix element is [22,23]

$$S_{fi} = \lim_{t\to\infty}\langle\phi_f(t)|\psi_i^+(t)\rangle, \qquad (2.14)$$

where the final state $\phi_f$ is a plane wave and $\psi_i$ is Eq. (2.7) above. Therefore, Eq. (2.14) is actually the project of (2.7) onto a plane wave in some azimuth angle. The concept of final state is different from that of (2.7).

The general scattering theory regarded the dashed lines in Fig. 1 as an intermediate state and one direction in the spherical wave as a plane wave.

What we want to emphasize is that the state before the scattering is the solid line and the final state is the dashed lines in Fig. 1. Equation (2.7) is the eigen state of Hamiltonian (2.5), while the state $\phi_f$ in (2.14) is not. Equation (2.8) is an approximate eigen state of the Hamiltonian.

In literature, when the fundamental scattering theory was employed, the one-particle Green's function of Schrödinger equation was resorted to, see Eq. (2.7). However, this theory was not applied to relativistic particles. One probable reason may be that no one explicitly put down the time-independent Green's functions for Dirac particles. They are now given in Table 4 in appendix A.

## 3 Evaluation of Differential Cross Section

In this section, we apply the fundamental theory introduced in the last section for all the three kinds of particles in three-dimensional (3D) space. We merely take the first order of Born approximation, i. e., Eq. (2.9). The same evaluation in one-dimensional (1D) and two-dimensional (2D) spaces is presented in appendix B.

### 3.1 Low-Momentum Motion

The procedure of treating elastic scattering was introduced in textbooks in detail. Here we just apply the formulas in the last section to show the standard routine that will be utilized uniformly to relativistic particles.

Let a free particle's Hamiltonian be

$$H_0 = -\frac{\hbar^2}{2m}\nabla^2. \tag{3.1}$$

Its continuous energy spectrum is

$$E_0 = \hbar^2 k^2 / 2m. \tag{3.2}$$

The initial state is a plane wave along the $z$ direction,

$$\varphi_0(\mathbf{r}) = C e^{i\mathbf{p}\cdot\mathbf{r}}, \mathbf{p} = p\mathbf{e}_z. \tag{3.3}$$

In this paper we use $\mathbf{e}_\gamma$ to represent the unit vector in the $\gamma$ direction.

The one-body Green's function corresponding to (3.1) is in the last row of Table 1 in appendix A.

$$G_0^+(\mathbf{r},\mathbf{r}';E) = -\frac{m}{2\pi\hbar^2}\frac{e^{ik|\mathbf{r}-\mathbf{r}'|}}{|\mathbf{r}-\mathbf{r}'|}, k = |\mathbf{k}| = |\mathbf{p}|. \tag{3.4}$$

Please notice that we merely consider elastic scattering. Substitution of (3.4) into (2.8) results in

$$\psi(\mathbf{r}) = \varphi_0(\mathbf{r}) - \int d\mathbf{r}' \frac{m}{2\pi\hbar^2}\frac{e^{ik|\mathbf{r}-\mathbf{r}'|}}{|\mathbf{r}-\mathbf{r}'|}H_1(\mathbf{r}')\varphi_0(\mathbf{r}'). \tag{3.5}$$

To a distant observer, the potential of scattering center can be neglected. In this case, the following approximations can be made for Eq. (3.5) [24,25]. In the exponent,

$$k|\mathbf{r}-\mathbf{r}'| \approx kr - k\mathbf{e}_r \cdot \mathbf{r}' = kr - \mathbf{k} \cdot \mathbf{r}', \qquad (3.6)$$

and in the denominator (3.5),

$$|\mathbf{r}-\mathbf{r}'| \approx |\mathbf{r}| = r. \qquad (3.7)$$

By these approximations, the second term of (3.5) is rewritten as

$$\varphi_s(\mathbf{r}) = -C \frac{e^{ikr}}{r} \frac{m}{2\pi\hbar^2} u. \qquad (3.8)$$

Here

$$u = \int d\mathbf{r}' e^{-i\mathbf{k}\cdot\mathbf{r}'} H_1(\mathbf{r}') e^{i\mathbf{p}\cdot\mathbf{r}'}. \qquad (3.9)$$

It is the transition matrix element of $H_1$ between two plane waves, but should not be regarded as the transition between the initial state $e^{i\mathbf{p}\cdot\mathbf{r}}$ and a final state $e^{i\mathbf{k}\cdot\mathbf{r}}$. The final state is (2.7), or, approximately, (2.8) instead of plane wave $e^{i\mathbf{k}\cdot\mathbf{r}'}$.

Comparing Eqs. (3.8) and (2.10), we obtain the expression of the scattering amplitude in this case,

$$f(\theta,\varphi) = -C \frac{m}{2\pi\hbar^2} u. \qquad (3.10)$$

The current density of low momentum particle is expressed by

$$\mathbf{j} = \frac{\hbar}{2im}(\psi^*\nabla\psi - \psi\nabla\psi^*). \qquad (3.11)$$

We substitute Eq. (3.5) into (3.11), and the resultant current density $\mathbf{j}$ includes indeed three terms just as (2.11). The first term is the current of the incident particle,

$$\mathbf{j}_0 = \frac{\hbar}{2im}(\varphi_0^*\nabla\varphi_0 - \varphi_0\nabla\varphi_0^*)\mathbf{e}_z = |C|^2 \frac{\hbar p}{m}\mathbf{e}_z. \qquad (3.12)$$

It remains in the $z$ direction, and does not contribute to any other direction.

For the other two terms in (2.11), we use the gradient operator in spherical coordinates.

$$\nabla = \mathbf{e}_r \frac{\partial}{\partial r} + \mathbf{e}_\theta \frac{1}{r}\frac{\partial}{\partial \theta} + \mathbf{e}_\varphi \frac{1}{r\sin\theta}\frac{\partial}{\partial \varphi}. \qquad (3.13)$$

The third term in (2.11) is the interference between the incident and scattered waves,

$$\mathbf{j}_c = \frac{\hbar}{2im}(\varphi_0^*\nabla\varphi_s - \varphi_0\nabla\varphi_s^* + \varphi_s^*\nabla\varphi_0 - \varphi_s\nabla\varphi_0^*). \qquad (3.14)$$

When Eqs. (3.3) and (3.8) are substituted into (3.14), it is seen that each term contains a factor like (2.12), so that can be neglected. Furthermore, if (3.13) is acted on (3.8), the derivatives with respect to angles will contribute a factor $1/r$, which makes the term proportional to $1/r^3$, a higher order small quantity [12]. As a result, $\mathbf{j}_c$ can be totally discarded.

The second term in (2.11) is the current density of scattered wave,

$$j_s = \frac{\hbar}{2im}(\varphi_s^* \nabla \varphi_s - \varphi_s \nabla \varphi_s^*). \tag{3.15}$$

We substitute (3.8) and (3.13) into (3.15) and keep terms proportional to $1/r^2$, dropping $1/r^3$ terms. In this process, it is seen that only the first term, $\nabla = e_r \frac{\partial}{\partial r}$ makes significant contribution. This shows that the $j_s$ is along the radial direction. The result is

$$j_s = \frac{\hbar k}{mr^2}|f|^2 e_r. \tag{3.16}$$

As last, substitution of Eqs. (3.12) and (3.16) into (2.13) yields the cross section,

$$\sigma(\theta,\varphi) = \frac{|f|^2}{|C|^2} = \frac{m^2}{4\pi^2\hbar^4}|u|^2. \tag{3.17}$$

A typical example is Coulomb potential scattering.

$$H_1 = \frac{q_1 q_2}{4\pi\varepsilon_0 r}. \tag{3.18}$$

When it is substituted into Eq. (3.9),

$$u = \frac{q_1 q_2}{4\pi\varepsilon_0} \frac{4\pi}{(\mathbf{k}-\mathbf{p})^2}. \tag{3.19}$$

In this case, the scattering amplitude (3.10) becomes

$$f(\theta,\varphi) = -\frac{Cmq_1q_2}{8\pi\varepsilon_0\hbar^2 k^2 \sin^2(\theta/2)}, \tag{3.20}$$

where

$$(\mathbf{k}-\mathbf{p})^2 = 4k^2 \sin^2\frac{\theta}{2}. \tag{3.21}$$

By Eq. (3.17), the cross section is

$$\sigma(\theta,\varphi) = \frac{(mq_1q_2)^2}{16(2\pi\varepsilon_0\hbar^2)^2 k^4 \sin^4(\theta/2)}. \tag{3.22}$$

This is Rutherford formula.

### 3.2 Relativistic Particles With Spin Zero

The author has pointed out [1] that for a free particle with positive kinetic energy, its Hamiltonian should be

$$H_0 = \sqrt{m^2c^4 - c^2\hbar^2\nabla^2}. \tag{3.23}$$

One application of this Hamiltonian was to figure out famous Klein paradox arising from when a relativistic particle encountered a square potential [2].

The eigen functions $\varphi_0$ of (3.23) is still Eq. (3.3), and its eigen energy is

$$E_k = \sqrt{m^2c^4 + c^2\hbar^2 k^2} \ . \tag{3.24}$$

The corresponding Green's function is in the last row of Table 2 in appendix A.

$$G_0^+ = \frac{E e^{ik|r-r'|}}{2\pi c^2 \hbar^2 |r-r'|}, k = \sqrt{E^2 - m^2c^4}/c\hbar \ . \tag{3.25}$$

When this Green's function is substituted into (2.7), we get the Lippmann-Schwinger equation for a relativistic particle with spin zero. The first order approximation (2.8) becomes

$$\psi(r) = \varphi_0 + \int dr' \frac{E}{2\pi c^2 \hbar^2} \frac{e^{ik|r-r'|}}{|r-r'|} H_1(r')\varphi_0(r') \ . \tag{3.26}$$

The second term is the $\varphi_s$ in (2.9). Equation (3.26) has the similar form as (3.5). Consequently, the approximations (3.6) and (3.7) also apply. Under these approximation, the second term in (3.26) is simplified to be the form of (3.8), and the expression of the scattering amplitude is

$$f(\theta,\varphi) = -C \frac{E}{2\pi c^2 \hbar^2} u \ , \tag{3.27}$$

where $u$ is Eq. (3.9).

Now let us calculate the current density $j(\psi)$ of the wave function (3.26). It has been pointed out that instead of (3.11), the expression of $j(\psi)$ should be as follows [1,26].

$$j_{(1)}(\psi) = \frac{i\hbar}{2m}(\psi^*\nabla\psi - \psi\nabla\psi^*), \tag{3.28a}$$

$$j_{(n)}(\psi) = \frac{i}{\hbar}\frac{\hbar^2}{2m} b_n \sum_{i=0}^{n-1} (-1)^i [\nabla^i \psi^* \nabla^{2n-i-1}\psi - \nabla^i \psi \nabla^{2n-i-1}\psi^*], n \geq 2, \tag{3.28b}$$

$$b_1 = 1, b_n = \frac{(2n-3)!!}{2^n n!}(\frac{\hbar^2}{m^2 c^2})^n, n \geq 2 \tag{3.28c}$$

and

$$j(\psi) = j_{(1)}(\psi) + j_{(2)}(\psi) + j_{(3)}(\psi) + \cdots = \sum_{i=1}^{\infty} j_{(i)}(\psi) \ . \tag{3.28d}$$

When (3.26) is substituted into (3.28), the resultant again contains three terms as (2.12). We discuss each term carefully.

Substitution of the $\varphi_0$ (3.3) into (3.28) results in

$$\boldsymbol{j}_0 = \boldsymbol{j}(\varphi_0) = \frac{c^2\hbar p}{E}\boldsymbol{e}_z. \tag{3.29}$$

It is along the incident direction and has no contribution to other directions.

The calculation of the interference current $\boldsymbol{j}_c$ is

$$\boldsymbol{j}_c = \boldsymbol{j}_{c(1)} + \boldsymbol{j}_{c(2)} + \boldsymbol{j}_{c(3)} + \cdots = \sum_{i=1}^{\infty} \boldsymbol{j}_{c(i)}, \tag{3.30a}$$

$$\boldsymbol{j}_{c(1)}(\psi) = \frac{i\hbar}{2m}(\varphi_0^*\nabla\varphi_s - \varphi_0\nabla\varphi_s^* + \varphi_s^*\nabla\varphi - \varphi_s\nabla\varphi_0^*) \tag{3.30b}$$

and

$$\boldsymbol{j}_{c(n)}(\psi) = \frac{i}{\hbar}\frac{\hbar^2}{2m}b_n\sum_{i=0}^{n-1}(-1)^i[\nabla^i\varphi_0^*\nabla^{2n-i-1}\varphi_s - \nabla^i\varphi_s\nabla^{2n-i-1}\varphi_0^* \\ +\nabla^i\varphi_s^*\nabla^{2n-i-1}\varphi_0 - \nabla^i\varphi_0\nabla^{2n-i-1}\varphi_s^*], n \geq 2. \tag{3.30c}$$

Now the $\varphi_0$ and $\varphi_s$ are (3.3) and (3.8), respectively. It is easily seen that each term of (3.30) contains a factor of (2.12) or its complex conjugate. The conclusion is that the $\boldsymbol{j}_c$ can be totally dropped. As for the current density of the scattered wave $\boldsymbol{j}_s = \boldsymbol{j}(\varphi_s)$, we substitute (3.13) and (3.8) which is of the form of $\varphi_s = \frac{e^{ikr}}{r}f$ into (3.28). Only the terms proportional to $1/r^2$ are retained, and those containing $1/r^3$ are dropped. So, only the first term in (3.13), $\nabla = \boldsymbol{e}_r\frac{\partial}{\partial r}$, actually acts. This shows that the current is along the radial direction. The result is that

$$\boldsymbol{j}_s(\varphi_s) = -\boldsymbol{e}_r\frac{imc^2}{\hbar}\sum_{n=1}^{\infty}\frac{(2n-3)!!}{2^n n!}(\frac{\hbar}{mc})^{2n}2n(ik)^{2n-1}\frac{|f|^2}{r^2} = \frac{|f|^2\hbar kc^2}{r^2 E}\boldsymbol{e}_r. \tag{3.31}$$

From Eqs. (3.29) and (3.31), the calculated cross section is

$$\sigma(\theta,\varphi) = r^2\frac{j_s}{j_0} = \frac{|f|^2}{|C|^2}. \tag{3.32}$$

This result is the same as that in textbooks [22,23]. However, we want to demonstrate some discrepancy between the procedures here and the textbooks.

In the present work, Eq. (3.28) is used to calculate the current density. When the wave function is the plane wave (3.3), the current is (3.29) which is relativistic velocity as it should be.

In the textbooks, only the first term of (3.28), i. e., Eq. (3.11), was used to calculate the current. When the wave function is the plane wave (3.3), the resultant was (3.12) which was the momentum divided by mass, not relativistic velocity. Usually, a normalization factor $\sqrt{mc^2/E}$ was attached to a relativistic wave function, see, e. g., Eq. (8.5) in [13]. This coefficient helped to manage the needed result of relativistic velocity. As a matter of fact, in evaluating the cross section, the normalization coefficient $C$ in (3.3) is not important in computing the cross section. Please notice that in Eq. (2.8), the $\varphi_0$ in the second term is just the first term. This is because Eq. (2.8) is resulted from iteration one time in Eq. (2.7). Hence, both terms in (2.9) have the same factor $C$. When calculating the cross section by (2.13), the $|C|^2$ in the numerator and denominator cancel out exactly, see Eqs. (3.17) and (3.32). That is why in this paper, we never explicitly put down the normalization coefficient $C$ in the $\varphi_0$.

### 3.3 Dirac Particles

The Hamiltonian of a free Dirac particle is

$$H_0 = c\boldsymbol{\alpha} \cdot \boldsymbol{p} + mc^2 \beta. \tag{3.33}$$

Its energy is

$$E_{(\pm)k} = \pm\sqrt{m^2c^4 + c^2\hbar^2 k^2}. \tag{3.34}$$

Every energy has two wave functions with spins up and down, respectively. Therefore, for every wave vector $\boldsymbol{k}$, there are four eigen wave functions. They are all plane waves. Here we pick up one with positive kinetic energy with spin up:

$$\varphi_0 = C\varphi_{(+)\uparrow} e^{i\boldsymbol{p}\cdot\boldsymbol{r}}, \tag{3.35a}$$

where

$$\varphi_{(+)\uparrow} = (1\ 0\ \zeta^{-1}\ 0)^+, \zeta = \sqrt{\frac{E+mc^2}{E-mc^2}}. \tag{3.35b}$$

The energy $E$ belongs to positive branch in Eq. (3.34). The corresponding Green's function is listed in the last row of Table 4 in appendix A.

$$G^+(\boldsymbol{r},\boldsymbol{r}';E>0) = \frac{e^{ik|\boldsymbol{r}-\boldsymbol{r}'|}}{|\boldsymbol{r}-\boldsymbol{r}'|}Q, \tag{3.36}$$

where

$$Q = \frac{k}{4\pi c\hbar}\begin{pmatrix} \zeta & [1+i/k|\boldsymbol{r}-\boldsymbol{r}'|]\boldsymbol{\sigma}\cdot\boldsymbol{e}_{\boldsymbol{r}-\boldsymbol{r}'} \\ [1+i/k|\boldsymbol{r}-\boldsymbol{r}'|]\boldsymbol{\sigma}\cdot\boldsymbol{e}_{\boldsymbol{r}-\boldsymbol{r}'} & 1/\zeta \end{pmatrix}. \tag{3.37}$$

Then Eq. (2.8) becomes

$$\psi(\boldsymbol{r}) = \varphi_0(\boldsymbol{r}) + \int d\boldsymbol{r}' \frac{e^{ik|\boldsymbol{r}-\boldsymbol{r}'|}}{|\boldsymbol{r}-\boldsymbol{r}'|} Q H_1(\boldsymbol{r}')\varphi_0(\boldsymbol{r}') = \varphi_0 + \varphi_s. \tag{3.38}$$

Again, the approximations (3.6) and (3.7) are made for this equation, and the second term is simplified into the form of (2.10). By the way, in Ref. [18] the asymptotic form (2.10) was also obtained, but the form of the Green's function there was too complicated to implement derivation. Hence, the result of Mott cross section was not presented there.

Let us consider a simplest case: $H_1$ is independent of spin and is a diagonal matrix. Subsequently, the scattering amplitude is expressed by

$$f(\theta,\varphi) = CQ\varphi_{(+)\uparrow}u, \tag{3.39}$$

where $u$ is still (3.9). Please note that the $\varphi_{(+)\uparrow}$ is a four-component spinor and the matrix $Q$ in (3.38) is of four order. Therefore, the $f(\theta,\varphi)$ is also a four-component spinor.

The current density of Dirac particles is evaluated by

$$\boldsymbol{j} = c\psi^{+}\boldsymbol{\alpha}\psi. \tag{3.40}$$

After substitution of (3.38) into (3.40), one gets the current again in the form of (2.11). Before calculating each term, we pay an attention to a fact that the direction of the vector $\boldsymbol{j}$, by virtue of the right hand side of (3.4), depends on the concrete form of the wave function. Let us retrospect that the expression (3.40) came from the continuity equation of Dirac particles: $\frac{\partial \rho}{\partial t} = -\nabla \cdot (c\psi^{+}\boldsymbol{\alpha}\psi)$. On the right hand side, the derivative with respect to the arguments of the function $\psi$ plays a role in determining the direction of $\boldsymbol{j}$. This is manifested in the following derivation.

For the incident particle, $\boldsymbol{j}_0 = c\varphi_0^{+}\boldsymbol{\alpha}\varphi_0$ and we must calculate $\nabla \cdot (c\varphi_0^{+}\boldsymbol{\alpha}\varphi_0)$. Because in (3.35) $\boldsymbol{p}$ is in the $z$ direction, only the $z$ component of $\boldsymbol{j}_0$ is nonzero,

$$\boldsymbol{j}_0 = c\varphi_0^{+}\alpha_z\varphi_0 \boldsymbol{e}_z = 2c|C|^2 \zeta^{-1}\boldsymbol{e}_z, \tag{3.41}$$

cf. Eq. (3.20) in [23]. In (3.41) here we do not explicitly put down the coefficient $C$.

The interference current is

$$\boldsymbol{j}_c = \varphi_0^{+}\boldsymbol{\alpha}\varphi_s + \varphi_s^{+}\boldsymbol{\alpha}\varphi_0. \tag{3.42}$$

Again, each term contains a factor of (2.12) and this interference current can be neglected.

The scattering current is

$$\boldsymbol{j}_s = \varphi_s^{+}\boldsymbol{\alpha}\varphi_s. \tag{3.43}$$

We use (3.13) to calculate $\nabla \cdot (\varphi_s^{+}\boldsymbol{\alpha}\varphi_s)$, and retain in $\varphi_s^{+}\boldsymbol{\alpha}\varphi_s$ the terms of order of

$1/r^2$, dropping those containing $1/r^3$. Hence, only the first term $\nabla = e_r \dfrac{\partial}{\partial r}$ in (3.13) is kept. As a result, the scattering current density is

$$j_s = |C|^2 \frac{1}{r^2} f^+ \alpha_r f e_r. \tag{3.44}$$

At this stage, we have to calculate the product of the Green's function (3.36) and incident wave function (3.35) comprised in the scattering amplitude in Eq. (3.39). This is a matrix product. It is noticed that in the exponent and denominator of (3.38), the dependences of $|\mathbf{r}-\mathbf{r}'|$ are the same as those in (3.5). This enables us to do the same approximations as (3.6) and (3.7). Meanwhile, in the off-diagonal elements in (3.37), $\mathrm{i}/k|\mathbf{r}-\mathbf{r}'| \approx \mathrm{i}/kr$ is regarded small compared to 1 so that can be neglected, and

$$\boldsymbol{\sigma} \cdot \boldsymbol{e}_{r-r'} \approx \boldsymbol{\sigma} \cdot \boldsymbol{e}_r = \sigma_r. \tag{3.45}$$

After these manipulations, the matrix products involved in (3.44) are carried out. The calculated scattering current density is

$$j_s = |C|^2 \frac{|u|^2}{r^2} \frac{E^2 \zeta^{-1}}{2\pi^2 c^3 \hbar^4} (1 - \frac{(c\hbar k)^2}{E^2} \sin^2 \frac{\theta}{2}) e_r. \tag{3.46}$$

Substituting Eqs. (3.41) and (3.46) into (2.13) results in

$$\sigma(\theta, \varphi) = \frac{E^2 |u|^2}{4\pi^2 c^4 \hbar^4}[1 - \frac{(c\hbar k)^2}{E^2} \sin^2 \frac{\theta}{2}]. \tag{3.47}$$

When $H_1$ is Coulomb potential, this result is famous Mott cross section [14,22,23].

We have assumed that $H_1$ is independent of spin, so that a particle subject to this potential will not flip its spin. Indeed, if we put down the wave function with spin down, $\varphi_{(+)\downarrow}$, it is easy to verify that

$$\varphi_{(+)\downarrow}^+ Q^+ \alpha_r Q \varphi_{(+)\uparrow} = 0. \tag{3.48}$$

The evaluation in [23] could not reflect this feature. Indeed, the matrix element for a right-handed relativistic electron to scatter to a left-handed one with interaction $\gamma^\mu$ was zero [22].

### 3.4 Discussions

We make some comparisons between the processes in this paper and in literature. In this work, the physical picture of scattering is clear, see Fig. 1. One-body Green's functions are given, which are needed in the scattering equation (2.7). Every approximation is explicitly given, e. g., Eqs. (3.6), (3.7) and (3.45). The derivation is concise. There is no need to do tedious traces of $\gamma$ matrices [22,23].

The key step in literature was to calculate the $S$ matrix element (2.14). The Green's functions were not employed. That the final state was appointed a plane wave actually made far-field approximation from the beginning. Therefore, one is difficult to clarify the approximations made in detail.

The definition of $u$ by (3.9) is the matrix element of $H_1$ between two plane waves. One plane wave is the initial state. The other is one with specific wave vector **k**. In the scattering amplitudes of all three kinds of particles, there is such a quantity, see (3.10), (3.27) and (3.39). It happens that this integral was always needed in evaluating the $S$ matrix element. For example, the $S$ matrix was just this integral, see page 194 in [13].

For all the three kinds of particles, the current density after scattering is composed of three terms as in Eq. (2.11), among which the interference current ***j***$_c$ can be discarded for it contains the oscillating factor shown by Eq. (2.12). In the general scattering theory, this term was never mentioned.

This work merely calculates the cross section up to the first order of Born approximation. The second and higher approximations will be dealt with later.

In literature, both the non-polarized and polarized scattering were considered [13,22,27]. The present work does not take into account the problem of polarization, which is left to be done later.

## 4 Summary

This work employs the fundamental scattering theory to evaluate the cross section up to the first order of Born approximation for low momentum particles and relativistic particles with spins 0 and 1/2. The cross section is expressed by the ratio of outgoing and ingoing current densities. The formulas are concise and uniform. Each kind of particles has its specific expressions of Green's function and current density.

The physical picture is clear. The one-body time-independent Green's functions are given. The approximations made in evaluation are elaborated in detail. This work does not resort to the concept of $S$ matrix.

For relativistic particles, our results are the same as those obtained by the general scattering theory in literature. However, we point out the discrepancies between the procedures in this work and in literature. In literature, for the relativistic particles with spin 0, the completeness relation of wave functions and the expression for current density were not proper; For Dirac particles, the final state was thought a plane wave, which was equivalent to taking a far-field approximation.

**Acknowledgements.** This work is supported by the National Key Research and Development Program of China [Grant No. 2018YFB0704304].

# Appendix

## A Free Particles' Green's Functions

In this appendix, we present the one-body time-independent Green's functions in one-, two- and three-dimensional spaces for all the three kinds of particles.

If a particle's Hamiltonian is *H*, the corresponding Green's function is defined as the solution satisfying the following equation and certain boundary conditions:

$$(z-H)G(\boldsymbol{r},\boldsymbol{r}';z) = \delta(\boldsymbol{r}-\boldsymbol{r}'),\tag{A.1}$$

where *z* is a complex parameter. This complex number is explicitly written as

$$z = E + \mathrm{i}b \tag{A.2}$$

where both *E* and *b* are real. *E* represents energy. If the eigenvalues $\{E_i\}$ and corresponding eigen functions $\{\psi_i\}$ of the *H* have been known, the Green's function can be evaluated by standard formula [15,16]

$$G(\boldsymbol{r},\boldsymbol{r}';z) = \sum_i \frac{\psi^*(\boldsymbol{r}')\psi(\boldsymbol{r})}{z-E_i}.\tag{A.3}$$

We consider free particles in the whole space. The boundary conditions are that the wave function is finite at infinity. Hence, the function has to be a plane wave with wave vector $\boldsymbol{k}$. The energy spectrum $E(\boldsymbol{k})$ is continuous. The summation in (A.2) is replaced by the integration with respect to wave vector $\boldsymbol{k}$.

$$G(\boldsymbol{r},\boldsymbol{r}';z) = \int \mathrm{d}\boldsymbol{k}\, \frac{\psi_{\boldsymbol{k}}^*(\boldsymbol{r}')\psi_{\boldsymbol{k}}(\boldsymbol{r})}{z-E(\boldsymbol{k})}.\tag{A.4}$$

It is well known that as the parameter *z* is just an eigenvalue, $z = E(\boldsymbol{k})$, the Green's function is not defined due to the zero in the denominator of (A.3). In this case, the side limits of the Green's function are defined as

$$G^{\pm}(\boldsymbol{r},\boldsymbol{r}';E) = G(\boldsymbol{r},\boldsymbol{r}';E\pm\mathrm{i}0^+).\tag{A.5}$$

In the following, the Green's functions for each kind of particles are presented.

### A.1 Low-Momentum Particles

Hamiltonian is

$$H = -\frac{\hbar^2}{2m}\nabla^2,\tag{A.6}$$

and its energy spectrum is

$$E_k = \frac{\hbar^2}{2m}k^2.\tag{A.7}$$

The corresponding Green's functions are listed in Table 1.

Table 1. The Green's functions and their side limits of low momentum particles. $H_0^{(1)}$ represents the first kind of Hankel function of the zero-th order.

| $z = E + \mathrm{i}b$ | $b > 0$ | $b < 0$ |
|---|---|---|
| 1D space | $G = \dfrac{m \mathrm{e}^{\mathrm{i}\sqrt{2mz}\|x-x'\|/\hbar}}{\mathrm{i}\hbar\sqrt{2mz}}$  $G^+ = \dfrac{m \mathrm{e}^{\mathrm{i}\sqrt{2mE}\|x-x'\|/\hbar}}{\mathrm{i}\hbar\sqrt{2mE}}$ | $G = \dfrac{\mathrm{i}m \mathrm{e}^{-\mathrm{i}\sqrt{2mz}\|x-x'\|/\hbar}}{2\hbar\sqrt{2mz}}$  $G^- = \dfrac{\mathrm{i}m \mathrm{e}^{-\mathrm{i}\sqrt{2mE}\|x-x'\|/\hbar}}{2\hbar\sqrt{2mE}}$ |
| 2D space | $G = \dfrac{mc}{2\mathrm{i}\hbar^2} H_0^{(1)}(\dfrac{\sqrt{2mz}}{\hbar}\|\mathbf{r}-\mathbf{r}'\|)$  $G^+ = \dfrac{mc}{2\mathrm{i}\hbar^2} H_0^{(1)}(\dfrac{\sqrt{2mE}}{\hbar}\|\mathbf{r}-\mathbf{r}'\|)$ | $G = \dfrac{mc}{2\mathrm{i}\hbar^2} H_0^{(1)}(-\dfrac{\sqrt{2mz}}{\hbar}\|\mathbf{r}-\mathbf{r}'\|)$  $G^- = \dfrac{mc}{2\mathrm{i}\hbar^2} H_0^{(1)}(-\dfrac{\sqrt{2mE}}{\hbar}\|\mathbf{r}-\mathbf{r}'\|)$ |
| 3D space | $G = \dfrac{m \mathrm{e}^{\mathrm{i}\sqrt{2mz/\hbar^2}\|\mathbf{r}-\mathbf{r}'\|}}{2\pi\hbar^2 \|\mathbf{r}-\mathbf{r}'\|}$  $G^+ = \dfrac{m \mathrm{e}^{\mathrm{i}\sqrt{2mE/\hbar^2}\|\mathbf{r}-\mathbf{r}'\|}}{2\pi\hbar^2 \|\mathbf{r}-\mathbf{r}'\|}$ | $G = \dfrac{m \mathrm{e}^{-\mathrm{i}\sqrt{2mz/\hbar^2}\|\mathbf{r}-\mathbf{r}'\|}}{2\pi\hbar^2 \|\mathbf{r}-\mathbf{r}'\|}$  $G^- = \dfrac{m \mathrm{e}^{-\mathrm{i}\sqrt{2mE/\hbar^2}\|\mathbf{r}-\mathbf{r}'\|}}{2\pi\hbar^2 \|\mathbf{r}-\mathbf{r}'\|}$ |

### A.2 Relativistic Particles With Spin Zero

A free particle's Hamiltonian is [1]

$$H_{(\pm)} = \pm\sqrt{m^2 c^4 - c^2 \hbar^2 \nabla^2} \tag{A.8}$$

and the energy spectrum is

$$E_{(\pm)\mathbf{k}} = \pm\sqrt{m^2 c^4 + c^2 \hbar^2 \mathbf{k}^2} . \tag{A.9}$$

The calculated Green's functions are listed in Table 2. For convenience we denote

$$k(z) = \sqrt{z^2 - m^2 c^4}/c\hbar . \tag{A.10}$$

Table 2. The Green's functions and their side limits of relativistic particles with spin zero.

| $z = E + ib$ | $(E > \sqrt{b^2 + m^2}, b > 0)$ $(E < -\sqrt{b^2 + m^2}, b < 0)$ | $(E > \sqrt{b^2 + m^2}, b < 0)$ $(E < -\sqrt{b^2 + m^2}, b > 0)$ |
|---|---|---|
| 1D space | $G = \dfrac{E e^{ik(z)|x-x'|}}{ic^2\hbar^2 k(z)}$ $G^+ = \dfrac{E e^{ik(E)|x-x'|}}{ic^2\hbar^2 k(E)}$ | $G = \dfrac{iE e^{-ik(z)|x-x'|}}{c^2\hbar^2 k(z)}$ $G^- = \dfrac{iE e^{-ik(E)|x-x'|}}{c^2\hbar^2 k(E)}$ |
| 2D space | $G = \dfrac{E}{2ic^2\hbar^2} H_0^{(1)}(k(z)|\mathbf{r}-\mathbf{r}'|)$ $G^+ = \dfrac{E}{2ic^2\hbar^2} H_0^{(1)}(-k(E)|\mathbf{r}-\mathbf{r}'|)$ | $G = \dfrac{E}{2ic^2\hbar^2} H_0^{(1)}(-k(z)|\mathbf{r}-\mathbf{r}'|)$ $G^- = \dfrac{E}{2ic^2\hbar^2} H_0^{(1)}(-k(E)|\mathbf{r}-\mathbf{r}'|)$ |
| 3D space | $G = \dfrac{E e^{ik(z)|\mathbf{r}-\mathbf{r}'|}}{2\pi c^2\hbar^2 |\mathbf{r}-\mathbf{r}'|}$ $G^+ = \dfrac{E e^{ik(E)|\mathbf{r}-\mathbf{r}'|}}{2\pi c^2\hbar^2 |\mathbf{r}-\mathbf{r}'|}$ | $G = \dfrac{E e^{-ik(z)|\mathbf{r}-\mathbf{r}'|}}{2\pi c^2\hbar^2 |\mathbf{r}-\mathbf{r}'|}$ $G^- = \dfrac{E e^{-ik(E)|\mathbf{r}-\mathbf{r}'|}}{2\pi c^2\hbar^2 |\mathbf{r}-\mathbf{r}'|}$ |

### A.3 Dirac Particles

The wave function of a Dirac particle is a multi-component spinor. Due to this reason, it is not easy to put down its time-independent Green's function in real space. It was of a simple form in momentum space [23], from which the form the real space can be obtained by Fourier transformation. Strange [18] tried to put down the expression of Dirac Green's function in the real 3D space, but it was rather complex. Up to now, no one has explicitly given the concrete and concise expression yet.

Fortunately, a method [18,28] was provided to express Dirac Green's function by means of Dirac Hamiltonian and low-momentum Green's function in 3D space. As a matter of fact, this method also applies to 1D and 2D spaces. Here we do this thing explicitly. First of all, we put down Dirac Hamiltonians.

$$H = c\sigma_1 p + mc^2 \sigma_3, \text{ 1D}. \tag{A.11}$$

$$H = c\boldsymbol{\sigma} \cdot \mathbf{p} + \sigma_3 mc^2, \boldsymbol{\sigma} = \sigma_1 \mathbf{e}_x + \sigma_2 \mathbf{e}_y, \text{ 2D} \tag{A.12}$$

$$H = c\boldsymbol{\alpha} \cdot \mathbf{p} + \beta mc^2, \text{ 3D}. \tag{A.13}$$

Here we have used Pauli matrices,

$$\sigma_1 = \begin{pmatrix} 0 & 1 \\ 1 & 0 \end{pmatrix}, \sigma_2 = \begin{pmatrix} 0 & -i \\ i & 0 \end{pmatrix}, \sigma_3 = \begin{pmatrix} 1 & 0 \\ 0 & -1 \end{pmatrix}. \tag{A.14}$$

The energy spectra are uniformly written as Eq. (A.9). One characteristic of all the three Hamiltonian is

$$(z+H)(z-H) = 2mc^2(Z + \frac{\hbar^2}{2m}\nabla^2), \tag{A.15}$$

where

$$Z = \frac{z^2 - m^2c^4}{2mc^2}. \tag{A.16}$$

We substitute the low-momentum Hamiltonian (A.6) into (A.1) and denote the low-momentum Green's function as $g$. Then $g$ satisfies the equation

$$(z + \frac{\hbar^2}{2m}\nabla^2)g(\mathbf{r},\mathbf{r}';z) = \delta(\mathbf{r}-\mathbf{r}'). \tag{A.17}$$

Now, in Eq. (A.17) we replace the parameter $z$ by $Z$ and make use of (A.15). The resultant is

$$\frac{1}{2mc^2}(z-H)(z+H)g(\mathbf{r},\mathbf{r}';Z) = \delta(\mathbf{r}-\mathbf{r}'). \tag{A.18}$$

Comparing Eqs. (A.18) and (A.1), one acquires the expression of Dirac Green's function as follows.

$$G(\mathbf{r},\mathbf{r}';z) = \frac{1}{2mc^2}(z+H)g(\mathbf{r},\mathbf{r}';Z). \tag{A.19}$$

In this way, Dirac particles' Green's functions can be written by means of low-momentum Green's functions. The latter have been listed in Table 1.

It is easily seen that under low-momentum approximation, the Green's function $G$ on the left hand side of (A.19) degrades to the $g$ on the right hand side. This is because when the momentum is very low, the off-diagonal matrix element of Dirac Hamiltonian can be neglected compared to diagonal ones. This means that

$$H \to mc^2 \begin{pmatrix} 1 & 0 \\ 0 & -1 \end{pmatrix} \text{ and } z+H \to \begin{pmatrix} z+mc^2 & 0 \\ 0 & z-mc^2 \end{pmatrix}. \tag{A.20}$$

As $z = mc^2 + z' \approx mc^2$, $Z = \frac{z^2 - m^2c^4}{2mc^2} \approx \frac{2mz'c^2}{2mc^2} = z'$. Then the upper diagonal matrix elements of the $G$ go back to the $g$.

$$G(\mathbf{r},\mathbf{r}';mc^2+z') \to g(\mathbf{r},\mathbf{r}';z'), \tag{A.21}$$

and the lower diagonal elements go to zero.

Please notice that there two parameters, $z$ and $Z$, in Eq. (A.19). Hence, the relationships between the real and imaginary arts of $z$ and $Z$ should be made clear. In

Table 3, we put down these relationships.

Table 3. Relationships between the real and imaginary arts of $z$ and $Z$ in Eq. (A.19), and the corresponding relationships between the side limits of the $G$ and $g$.

| $z = E + ib$ | $b > 0$ | $b < 0$ |
|---|---|---|
| $E > \sqrt{b^2 + m^2 c^4}$ | $\text{Re } Z > 0, \text{Im } Z > 0$ | $\text{Re } Z > 0, \text{Im } Z < 0$ |
| $G(\boldsymbol{r}, \boldsymbol{r}'; z)$ | $g(\boldsymbol{r}, \boldsymbol{r}'; Z)$ | $g(\boldsymbol{r}, \boldsymbol{r}'; Z)$ |
| $G^+(\boldsymbol{r}, \boldsymbol{r}'; E)$ | $g^+(\boldsymbol{r}, \boldsymbol{r}'; \text{Re } Z)$ | $g^-(\boldsymbol{r}, \boldsymbol{r}'; \text{Re } Z)$ |
| $E < -\sqrt{b^2 + m^2 c^4}$ | $\text{Re } Z > 0, \text{Im } Z < 0$ | $\text{Re } Z > 0, \text{Im } Z > 0$ |
| $G(\boldsymbol{r}, \boldsymbol{r}'; z)$ | $g(\boldsymbol{r}, \boldsymbol{r}'; Z)$ | $g(\boldsymbol{r}, \boldsymbol{r}'; Z)$ |
| $G^-(\boldsymbol{r}, \boldsymbol{r}'; E)$ | $g^-(\boldsymbol{r}, \boldsymbol{r}'; \text{Re } Z)$ | $g^+(\boldsymbol{r}, \boldsymbol{r}'; \text{Re } Z)$ |

With the above preparation, we are now able to write the concrete expression of time-independent Dirac Green's in 1D, 2D and 3D space. They are listed in Table 4. For the sake of convenience, the following denotations are made.

$$P_1(z) = \begin{pmatrix} \zeta(z) & \text{sgn}(x - x') \\ \text{sgn}(x - x') & \zeta^{-1}(z) \end{pmatrix}. \tag{A.22}$$

$$\zeta(z) = \frac{\sqrt{z + mc^2}}{\sqrt{z - mc^2}}. \tag{A.23}$$

$$P_{2\pm}(z) = \begin{pmatrix} \zeta(z) H_0^{(1)} & \pm \frac{i(x - x') - (y - y)}{|\boldsymbol{r} - \boldsymbol{r}'|} H_1^{(1)} \\ \pm \frac{i(x - x') + (y - y)}{|\boldsymbol{r} - \boldsymbol{r}'|} H_1^{(1)} & \zeta^{-1}(z) H_0^{(1)} \end{pmatrix}. \tag{A.24}$$

Here $H_1^{(1)}$ is the first order Hankel function of the first kind, and the argument of $H_0^{(1)}$ and $H_1^{(1)}$ is $k(z)|\boldsymbol{r} - \boldsymbol{r}'|$.

$$P_{3\pm}(z) = \begin{pmatrix} \zeta(z) & [\pm 1 + \frac{i}{k(z)|\boldsymbol{r} - \boldsymbol{r}'|}] \boldsymbol{\sigma} \cdot \boldsymbol{e}_{r-r'} \\ [\pm 1 + \frac{i}{k(z)|\boldsymbol{r} - \boldsymbol{r}'|}] \boldsymbol{\sigma} \cdot \boldsymbol{e}_{r-r'} & \zeta^{-1}(z) \end{pmatrix}. \tag{A.25}$$

Table 4. Time-independent Green's functions of free Dirac particles and their side limits.

| $z = E + \mathrm{i}b$ | $(E > \sqrt{b^2 + m^2}, b > 0)$ $(E < -\sqrt{b^2 + m^2}, b < 0)$ | $(E > \sqrt{b^2 + m^2}, b < 0)$ $(E < -\sqrt{b^2 + m^2}, b > 0)$ |
|---|---|---|
| 1D space | $G = \dfrac{1}{2\mathrm{i}c\hbar^2} \mathrm{e}^{\mathrm{i}k(z)|x-x'|} P_1(z)$ $G^+ = \dfrac{1}{2\mathrm{i}c\hbar} \mathrm{e}^{\mathrm{i}k(E)|x-x'|} P_1(E)$ | $G = \dfrac{\mathrm{i}}{2c\hbar} \mathrm{e}^{-\mathrm{i}k(z)|x-x'|} P_1(z)$ $G^- = \dfrac{\mathrm{i}}{2c\hbar} \mathrm{e}^{-\mathrm{i}k(E)|x-x'|} P_1(E)$ |
| 2D space | $G = \dfrac{k(z)}{4\mathrm{i}c\hbar} P_{2+}(z)$ $G^+ = \dfrac{k(E)}{4\mathrm{i}c\hbar} P_{2+}(E)$ | $G = \dfrac{k(z)}{4\mathrm{i}c\hbar} P_{2-}(-z)$ $G^- = \dfrac{k(E)}{4\mathrm{i}c\hbar} P_{2-}(-E)$ |
| 3D space | $G = \dfrac{k(z)\mathrm{e}^{\mathrm{i}k(z)|\mathbf{r}-\mathbf{r}'|}}{4\pi c\hbar |\mathbf{r}-\mathbf{r}'|} P_{3+}(z)$ $G^+ = \dfrac{k(E)\mathrm{e}^{\mathrm{i}k(E)|\mathbf{r}-\mathbf{r}'|}}{4\pi c\hbar |\mathbf{r}-\mathbf{r}'|} P_{3+}(E)$ | $G = \dfrac{k(z)\mathrm{e}^{-\mathrm{i}k(z)|\mathbf{r}-\mathbf{r}'|}}{4\pi c\hbar |\mathbf{r}-\mathbf{r}'|} P_{3-}(z)$ $G^- = \dfrac{k(E)\mathrm{e}^{-\mathrm{i}k(E)|\mathbf{r}-\mathbf{r}'|}}{4\pi c\hbar |\mathbf{r}-\mathbf{r}'|} P_{3-}(E)$ |

One point should be supplemented. The 1D and 2D Dirac Hamiltonians are (A.11) and (A.12), respectively. Each has two solutions. The 3D Dirac Hamiltonian (A.13) has four solutions. The summation in Eq. (A.3) should comprise all solutions of a Hamiltonian. As an example, the four solutions in 3D case are denoted by $\psi^+_{(+)\uparrow}$, $\psi^+_{(+)\downarrow}$, $\psi^+_{(-)\uparrow}$ and $\psi^+_{(-)\downarrow}$, respectively, all being column spinors. Then Eq. (A.4) should be explicitly written as

$$G(\mathbf{r},\mathbf{r}';z) = \int \mathrm{d}\mathbf{k} \frac{\psi^+_{(+)\uparrow \mathbf{k}}(\mathbf{r}')\psi_{(+)\uparrow \mathbf{k}}(\mathbf{r})}{z - E_{(+)\mathbf{k}}} + \int \mathrm{d}\mathbf{k} \frac{\psi^+_{(+)\downarrow \mathbf{k}}(\mathbf{r}')\psi_{(+)\downarrow \mathbf{k}}(\mathbf{r})}{z - E_{(+)\mathbf{k}}}$$
$$+ \int \mathrm{d}\mathbf{k} \frac{\psi^+_{(-)\uparrow \mathbf{k}}(\mathbf{r}')\psi_{(-)\uparrow \mathbf{k}}(\mathbf{r})}{z - E_{(-)\mathbf{k}}} + \int \mathrm{d}\mathbf{k} \frac{\psi^+_{(-)\downarrow \mathbf{k}}(\mathbf{r}')\psi_{(-)\downarrow \mathbf{k}}(\mathbf{r})}{z - E_{(-)\mathbf{k}}}.$$

(A.26)

There are four terms altogether. Among them, two with positive energy are the same, and two with negative energy are the same. In Table 4, only one term with positive or negative energy is given.

## B. Elastic Scattering in One- and Two-Dimensional Spaces

In the main text, we evaluated the cross sections of elastic scattering in 3D space. In this appendix, we give the results in 1D and 2D spaces, showing that they are produced through the same procedure.

### B.1 One-dimensional space

The current densities of the incident and scattered waves are denoted by $j_0$ and $j_s$, respectively. The cross section is defined by

$$\sigma = j_s / j_0. \tag{B.1}$$

Please note that it differs from (2.13) for 3D space. The scattering equation (2.8) is simplified into 1D form:

$$\psi(x) = \varphi_0(x) + \int dx' G_0^+(x, x'; E_i) H_1(x') \varphi_0(x'). \tag{B.2}$$

The incident wave is along positive $x$ direction, and the scattered wave $\varphi_s$ has merely two directions: positive and negative $x$ directions.

**1. low-momentum particles**

The incident wave is

$$\varphi_0(x) = C e^{ipx}. \tag{B.3}$$

The Green's function was given in the second row of Table 1.

$$G^+(x, x'; E) = \frac{m}{i\hbar^2 k} e^{ik|x-x'|}, k = \frac{\sqrt{2mE}}{\hbar}. \tag{B.4}$$

As $x \gg x'$, the wave function after scattering can be approximated to be

$$\psi(x) = \varphi_0(x) + e^{ikx} f, k = p, \tag{B.5}$$

where scattering amplitude is

$$f = C \frac{m}{i\hbar^2 k} u_1. \tag{B.6}$$

Here we have defined

$$u_1 = \int dx' e^{-ikx'} H_1(x') e^{ipx'}. \tag{B.7}$$

It is of the same form of (3.9) in 3D space. Substituting (B.5) into the expression current density $j = \frac{\hbar}{2im}(\psi^* \psi' - \psi \psi^{*'})$, we obtain

$$j = j_0 + j_s + j_c. \tag{B.8}$$

The incident current density is

$$j_0 = \frac{\hbar}{2im}(\varphi_0^* \varphi_0' - \varphi_0^{*\prime}\varphi_0) = |C|^2 \frac{\hbar p}{m}. \tag{B.9}$$

The interfere current density $j_c$ is again neglected. The scattering current density is

$$j_s = \frac{\hbar}{2im}(\varphi_s^* \varphi_s' - \varphi_s^{*\prime}\varphi_s) = \frac{\hbar k}{m}|f|^2. \tag{B.10}$$

The cross section is calculated by Eqs. (B.9), (B.10) and (B.1),

$$\sigma = \frac{|f|^2}{|C|^2} = \frac{m^2}{\hbar^4 k^2}|u_1|^2. \tag{B.11}$$

### 2. Relativistic particles with spin zero

The scattering equation is still (B.2), and the incident wave is (B.3). The Green's function was given in the second row of Table 2.

$$G^+ = \frac{E e^{ik|x-x'|}}{ic^2 \hbar^2 k}, \quad k = \frac{\sqrt{E^2 - m^2 c^4}}{c\hbar}. \tag{B.12}$$

This Green's function is substituted into Eq. (B.2) and we have

$$\psi(x) = \varphi_0(x) + \int dx' \frac{E}{ic^2 \hbar^2 k} e^{ik|x-x'|} H_1(x')\varphi_0(x') = \varphi_0(x) + e^{ikx}f, \tag{B.13}$$

where in the latter step the approximation for $x \gg x'$ has been made. This expression is the same as (B.5), and the scattering amplitude $f$ is (B.6). The wave function (B.13) is substituted into (3.28) to calculate the current density. There are again three terms as in Eq. (B.8). The interference current $j_c$ is discarded. The current densities of the incident and scattered waves are respectively

$$j_0 = |C|^2 \frac{\hbar k c^2}{E} \quad \text{and} \quad j_s = \frac{\hbar k c^2}{E}|f|^2. \tag{B.14}$$

It results from (B.1) that the cross section is

$$\sigma = \frac{j_s}{j_0} = \frac{E^2}{c^4 \hbar^4 k^2}|u_1|^2. \tag{B.15}$$

In the low-momentum limit, $E \to mc^2$, Eq. (B.15) tends to be (B.11).

### 3. Dirac particles

The scattering equation is still (B.2). The plane wave with positive energy is chosen as the incident wave,

$$\varphi_{0(+)} = C\varphi_{(+)} e^{ipx/\hbar}, \varphi_{(+)} = \begin{pmatrix} 1 \\ \zeta^{-1}(E) \end{pmatrix}, \tag{B.16}$$

where $\zeta$ is (A.23). The Green's function has been given in the second row of Table 4.

$$G^+ = \frac{1}{2ic\hbar} e^{ik(E)|x-x'|} P_1(E).  \tag{B.17}$$

It is used in Eq. (B.2), and then the approximation for $x \gg x'$ is made. We obtain the form of Eq. (B.5). The scattering amplitude is

$$f = CQ\varphi_{0(+)}u_1, \tag{B.18}$$

where $u_1$ is still Eq. (B.7) and

$$Q = \frac{1}{2ic\hbar}\begin{pmatrix} \zeta(E) & 1 \\ 1 & \zeta^{-1}(E) \end{pmatrix}. \tag{B.19}$$

The function $\psi$ in the form of Eq. (B.5) is substituted into the expression of current density $j = \psi^+\sigma_1\psi$. There are again three terms as in Eq. (B.8). At last, the calculated cross section is

$$\sigma = \frac{j_s}{j_0} = \frac{\varphi_{(+)}^+ Q^+ \sigma_1 Q \varphi_{(+)} |u_1|^2}{\varphi_{(+)}^+ \sigma_1 \varphi_{(+)}} = \frac{E^2}{c^4\hbar^4 k^2}|u_1|^2. \tag{B.20}$$

This is the same as Eq. (B.15) for relativistic particles with spin zero. The reason is that in 1D space there is no spin.

### B.2 Two-Dimensional Space

The polar coordinates in 2D space are $(r,\theta)$. The cross section is defined by

$$\sigma(\theta) = rj_s / j_0. \tag{B.21}$$

The scattering equation is again in the form of (2.8),

$$\psi(\boldsymbol{r}) = \varphi_0(\boldsymbol{r}) + \int d\boldsymbol{r}' G_0^+(\boldsymbol{r},\boldsymbol{r}';E) H_1(\boldsymbol{r}')\varphi_0(\boldsymbol{r}'). \tag{B.22}$$

**1. low-momentum particles**
The incident wave is

$$\varphi_0(\boldsymbol{r}) = Ce^{i\boldsymbol{p}\cdot\boldsymbol{r}}. \tag{B.23}$$

The Green's function is seen in the third row of Table 1.

$$G^+(\boldsymbol{r},\boldsymbol{r}';E) = \frac{m}{2i\hbar^2} H_0^{(1)}(k|\boldsymbol{r}-\boldsymbol{r}'|), k = \frac{\sqrt{2mE}}{\hbar}. \tag{B.24}$$

As $|\boldsymbol{r}-\boldsymbol{r}'|$ is very large, we take the asymptotic form of Hankel function $H_0^{(1)}$. Then Eq. (B.22) is approximated to be

$$\psi(\boldsymbol{r}) = \varphi_0(\boldsymbol{r}) + \frac{e^{ikr}}{\sqrt{r}} f, \tag{B.25}$$

where the scattering amplitude is

$$f = C\frac{me^{-i\pi/4}}{i\hbar^2\sqrt{2\pi k}}u_2. \tag{B.26}$$

Here we have defined

$$u_2 = \int d\mathbf{r}' e^{-i\mathbf{k}\cdot\mathbf{r}'}H_1(\mathbf{r}')\varphi_0(\mathbf{r}')e^{i\mathbf{p}\cdot\mathbf{r}'}. \tag{B.27}$$

When Eq. (B.25) is substituted into the expression of current density $\mathbf{j} = \frac{\hbar}{2im}(\psi^*\nabla\psi - \psi\nabla\psi^*)$, we again obtain three terms as in Eq. (2.13). The interference current $\mathbf{j}_c$ is dropped. In calculating the scattering current $\mathbf{j}_s$, we keep the terms proportional to $1/r$, neglecting the terms containing $1/r^2$. The result is that

$$\mathbf{j}_s = \frac{\hbar k |f|^2}{mr}\mathbf{e}_r. \tag{B.28}$$

The calculated cross section is

$$\sigma(\theta) = r\frac{j_s}{j_0} = \frac{|f|^2}{|C|^2} = \frac{m^2}{2\pi\hbar^4 k}|u_2|^2. \tag{B.29}$$

In Ref. [29], the scattering amplitude in 2D space was expanded by partial waves.

**2. Relativistic particles with spin zero**

The scattering equation is still (B.22). The incident wave is (B.23). Green's function is listed in the third row of Table 2.

$$G^+ = \frac{E}{2ic^2\hbar^2}H_0^{(1)}(k|\mathbf{r}-\mathbf{r}'|), k = \frac{\sqrt{E^2 - m^2c^4}}{c\hbar}. \tag{B.30}$$

As $|\mathbf{r}-\mathbf{r}'|$ is very large, we take the asymptotic form of $H_0^{(1)}$. Then Eq. (B.22) is approximated to be

$$\psi(\mathbf{r}) = \varphi_0(\mathbf{r}) + \frac{e^{ikr}}{\sqrt{r}}f. \tag{B.31}$$

This form is the same as Eq. (B.25) but with

$$f = C\frac{Ee^{-i\pi/4}}{ic^2\hbar^2\sqrt{2\pi k}}u_2. \tag{B.32}$$

The wave function (B.31) is substituted into Eq. (3.28). The current comprises three terms as in Eq. (2.13). The interference current is discarded. In calculating scattering current, only the terms proportional to $1/r$ are retained, and those containing $1/r^2$ dropped. The calculated cross section is

$$\sigma(\theta) = r\frac{j_s}{j_0} = \frac{E^2}{2\pi c^4 \hbar^4 k}|u_2|^2. \tag{B.33}$$

In the low-momentum limit, $E \to mc^2$, it goes to Eq. (B.29).

### 3. Dirac particles

The scattering equation is again (B.22). A plane wave with positive energy is chosen as the incident wave.

$$\varphi_{0(+)} = C\varphi_{(+)}e^{i\mathbf{p}\cdot\mathbf{r}}, \varphi_{(+)} = \begin{pmatrix} c\hbar(p_x + ip_y) \\ E_{(+)} - mc^2 \end{pmatrix}. \tag{B.34}$$

The Green's function was listed in the third row of Table 4.

$$G^+ = \frac{k(E)}{4ic\hbar}P_{2+}(E). \tag{B.35}$$

As $|\mathbf{r}| \gg |\mathbf{r}'|$, we take the asymptotic form of Hankel function. Then the Green's function is approximated to be

$$G^+(\mathbf{r},\mathbf{r}';E) = \frac{e^{ikr}}{\sqrt{r}}Qe^{-i\mathbf{k}\cdot\mathbf{r}'}, \tag{B.36}$$

where

$$Q = \frac{1}{4ic\hbar}\sqrt{\frac{2k}{\pi r}}\begin{pmatrix} \zeta & (x+iy)/r \\ (x-iy)/r & \zeta^{-1} \end{pmatrix}. \tag{B.37}$$

The wave function after scattering is in the form of Eq. (B.25), where the scattering amplitude is

$$f = CQ\varphi_{0(+)}u_2. \tag{B.38}$$

Here the $u_2$ is Eq. (B.27). Now the current density is calculated by

$$\mathbf{j} = \psi^+ \begin{pmatrix} 0 & \mathbf{e}_x + i\mathbf{e}_y \\ \mathbf{e}_x - i\mathbf{e}_y & 0 \end{pmatrix}\psi. \tag{B.39}$$

The wave function in the form of (B.25) is substituted into (B.39). There are three terms as in Eq. (2.13). Finally, the calculated cross section is

$$\sigma(\theta) = r\frac{j_s}{j_0} = \frac{E^2}{2\pi c^4 \hbar^4 k}|u_2|^2. \tag{B.40}$$

This result is the same as (B.33), showing that in 2D space, a particle does not manifest spin in elastic scattering.

**References**

[1] Huai-Yu Wang, New results by low momentum approximation from relativistic quantum mechanics equations and suggestion of experiments, *J. Phys. Commun.*, **4**: (2020), 125004. https://dx.doi.org/10.1088/2399-6528/abd00b

[2] Huai-Yu Wang, Solving Klein's paradox, *J. Phys. Commun.*, **4** (2020), 125010. https://doi.org/10.1088/2399-6528/abd340

[3] Huai-Yu Wang, Fundamental formalism of statistical mechanics and thermodynamics of negative kinetic energy systems, *J. Phys. Commun.*, **5** (2021), 055012. https://doi.org/10.1088/2399-6528/abfe71

[4] Huai-Yu Wang, Macromechanics and two-body problems, *J. Phys. Commun.*, **5** (2021), 055018. http://dx.doi.org/10.4006/0836-1398-35.2.152

[5] Huai-Yu Wang, The modified fundamental equations of quantum mechanics, *Physics Essays*, **35**(2) (2022), 152-164. http://dx.doi.org/10.4006/0836-1398-35.2.152

[6] Huai-Yu Wang, Virial theorem and its symmetry, *Journal of North China Institute of Science and Technology*, **18**(4) (2021), 1-10. (in Chinese)
DOI: http://19956/j.cnki.ncist.2021.04.001

[7] Huai-Yu Wang, Resolving problems of one-dimensional potential barriers based on Dirac equation, *Journal of North China Institute of Science and Technology*, **19**(1): (2022), 97-107. (in Chinese) DOI: http://10.19956/j.cnki.ncist.2022.01.016

[8] L. S. Rodberg and R. M. Thaler, 1967 Introduction to the Quantum theory of Scattering, (New York: Academic Press), (1967).

[9] L. D. Landau and E. M. Lifshitz, Chap.7 in Quantum Mechanics Non-relativistic Theory, Vol. 3 of course of Theoretical Physics, (New York: Pergmon Press), (1977).

[10] A. Messiah, Chap. 10 in Quantum Mechanics, I 2$^{nd}$ ed. (Mineloa, New York: Dover Publications), (2014).

[11] B. A. Lippmann and J. Schwinger, Variational Principles for Scattering Processes. I, *Phys. Rev.* **79** (1950), 469. https://doi.org/10.1103/PhysRev.79.469

[12] S. Weinberg, Chaps. 7 and 8 in Lectures on Quantum Mechanics, 2$^{nd}$ ed. Cambridge: Cambridge University Press), (2015).

[13] A. Wachter, Chap. 7 in Relativistic Quantum Mechanics, (Springer Science+Business Media B.V.), (2011). DOI 10.1007/978-90-481-3645-2

[14] N. F. Mott, The Scattering of Fast Electrons by Atomic Nuclei, *Proceedings of the Royal Society of London. Series A*, Containing Papers of a Mathematical and Physical Character **124**(794) (1929), 425-442.
Stable URL: https://www.jstor.org/stable/95377

[15] Huai-Yu Wang, Chap. 3 in Green's Function in Condensed Matter Physics, (Beijing and Oxford: Science Press and Alpha Science International Ltd.), (2012).

[16] Huai-Yu Wang, Chap. 6 in Mathematics for Physicists, (Singapore: Science Press and World Scientific), (2017).

[17] J. J. Sakura, Chap. 7 in Modern Quantum Mechanics, (Massachusetts: Addison-

Wesley Publishing Company, Inc.), (1994).
[18] P. Strange, Chap. 11 in Relativistic Quantum Mechanics, (Cambridge: Cambridge University Press), (1998).
[19] S. Flügge, Practical Quantum Mechanics, Vol. I (Berlin: Springer-Verlag), (1999), 208-210.
[20] R. G. Newton, Scattering Theory of Waves and Particles, (New York: Spring-Verlag New York Inc.), (1982).
[21] L. E. Ballentine, Chap. 16 in Quantum Mechanics A Modern Development, 2$^{nd}$ ed. (Singapore: World Scientific), (2015).
[22] J. D. Bjorken and S. D. Drell, Relativistic Quantum Mechanics, (New York: McGraw-Hill Book Company), (1964).
[23] W. Greiner and J. Reinhardt, Chaps. 1-3 and 8 in Quantum Electrodynamics, 4$^{th}$ed. (Berlin: Springer-Verlag), (2009).
[24] L. I. Schiff, Chap. 9 in Quantum Mechanics, 3$^{rd}$ ed., (New York: McGraw Hill Book Company), (1968).
[25] A. Messiah, Chap. 19 in Mecanique Quantique II, (Paris: Dunod), (1973).
[26] K. Kowalski and J. Rembieli´nski, Salpeter equation and probability current in the relativistic Hamiltonian quantum mechanics, *Phys. Rev.* A **84** (2011), 012108. DOI: https://doi.org/10.1103/PhysRevA.84.012108
[27] H. A. Tolhokk, Electron polarization, theory and experiment, *Rev. Mod. Phys.*, **28**(3) (1956), 277. DOI: https://doi.org/10.1103/RevModPhys.28.277
[28] J. Blackledge and B. Babajanov, On the Dirac Scattering Problem, *Mathematica Aeterna* **3** (2013), 535. doi:10.21427/D7JP6B
[29] G. D. Mahan, Quantum Mechanics in a nutshell, (Princeton: Princeton University Press), (2005), 326-328.